\documentclass[twocolumn,longbibliography, superscriptaddress, english, pre]{revtex4-1}
\usepackage[colorlinks=true,urlcolor=blue,citecolor=blue,linkcolor=blue]{hyperref}
\usepackage[T1]{fontenc}
\usepackage[latin9]{inputenc}
\usepackage{amssymb}
\usepackage{graphicx}
\usepackage{amsmath,color}
\usepackage{mathrsfs}
\usepackage{float}
\usepackage{indentfirst}
\usepackage{braket}
\usepackage{comment}
\usepackage{txfonts}

\makeatletter

\def\journal #1, #2, #3, 1#4#5#6{{\sl #1~}{\bf #2}, #3 (1#4#5#6) }

\def\eqa{\begin{eqnarray}}
\def\eea{\end{eqnarray}}
\newcommand{\eq}{\begin{equation}}
\newcommand{\ee}{\end{equation}}

\newcommand{\Eq}[1]{Eq.~(\ref{#1})}

\newcommand{\Tr}{{\rm Tr}}

\renewcommand{\mathbf}[1]{\ensuremath{\boldsymbol{ #1}}  }

\makeatother

\usepackage{babel}

\begin{document}

\title{Information Perspective  to Probabilistic Modeling:  Boltzmann Machines versus Born Machines}

\author{Song Cheng}
\affiliation{Institute of Physics, Chinese Academy of Sciences, Beijing 100190, China}
\affiliation{University of Chinese Academy of Sciences, Beijing, 100049, China}
\author{Jing Chen}
\affiliation{Institute of Physics, Chinese Academy of Sciences, Beijing 100190, China}
\affiliation{University of Chinese Academy of Sciences, Beijing, 100049, China}
\author{Lei Wang}
\affiliation{Institute of Physics, Chinese Academy of Sciences, Beijing 100190, China}

\begin{abstract}
We compare and contrast the statistical physics and quantum physics inspired approaches for unsupervised generative modeling of classical data. The two approaches represent probabilities of observed data using energy-based models and quantum states respectively. 
Classical and quantum information patterns of the target datasets therefore provide principled guidelines for structural design and learning in these two approaches. Taking the restricted Boltzmann machines (RBM) as an example, we analyze the information theoretical bounds of the two approaches. We verify our reasonings by comparing the performance of RBMs of various architectures on the standard MNIST datasets. 
\end{abstract}
\maketitle

\section{Introduction}
The fruitful interplay between statistical physics and machine learning dates back to at least the early studies of spin glasses and neural networks~\cite{Hopfield1982, PhysRevA.32.1007}. The two fields share common interests on emergent collective behavior of complex systems with a large number of degrees of freedom. In particular, unsupervised generative modeling is closely related to the inverse statistical problems~\cite{doi:10.1080/00018732.2017.1341604}, where one infers parameters of a model based on observations. The model can generate new samples according to the learned probability distribution, hence the name generative modeling.   
Inspired by the statistical physics, one can model the data probability according to the Boltzmann distribution with an energy function of the observed variables
\begin{equation}
p(\mathbf{v}) = \frac{e^{-E(\mathbf{v})}}{Z}, 
\label{eq:boltzmannmachine}
\end{equation}
where $Z=\sum_{\mathbf{v}} e^{-E(\mathbf{v})}$, the partition function, is the normalization factor of the probability density. The functional form of the energy function is typically predetermined to deliver certain prior knowledge about the data. Structured probabilistic models of the form (\ref{eq:boltzmannmachine}) are collectively denoted as energy-based models~\cite{Goodfellow-et-al-2016-Book}, in which the prominent examples are the \emph{Boltzmann Machine}~\cite{hinton1986learning}.

On the other hand, by exploiting the inherent probabilistic nature of quantum mechanics, one can model the probability distribution using a quantum state 
\begin{equation}
p(\mathbf{v}) = \frac{|\Psi(\mathbf{v})|^{2}}{N},
\label{eq:bornmachine}
\end{equation}
where $N=\sum_{\mathbf{v}}|\Psi(\mathbf{v})|^{2} $ is the normalization factor. The square ensures the positivity of the probability. Recently, in conjunction with the applications of machine learning techniques to quantum physics problems~\cite{Carleo:2016vp, 2017arXiv170104844D, 2017arXiv170105039G, Huang2017, 2017arXiv170305334T, 2017arXiv170405148C, Clark2017, Glasser2017, Kaubruegger2017}, there emerges a quantum perspective to problems in machine learning~\cite{MPSSL, RBMvsTN, 1704.01552, 2017arXiv170901662H, Liu2017, 2017arXiv171005520Z, 2017arXiv171010248P, 1711.02038, 2017arXiv171104606H}. In particular, Equation (\ref{eq:bornmachine}) translates the generative modeling of probability density to the problem of learning a quantum state. In fact, the necessity of this quantum interpretation was also anticipated in earlier machine learning literature. The mathematical structure of quantum mechanics appears naturally when one explores more flexible models than \Eq{eq:boltzmannmachine} while still attempts to ensure the positivity of the probability density~\cite{pmlr-v20-bailly11, Zhao-Jaeger}. We call these approaches \emph{Born Machines} to acknowledge the probabilistic interpretation of the quantum mechanics~\cite{MaxBorn}. 

Both Eqs.~(\ref{eq:boltzmannmachine}) and ~(\ref{eq:bornmachine}) allow one to import insights and intuitions from statistical and quantum physics to unsupervised generative modeling. Physical considerations can be used to assess the complexity of the dataset and the representational power of the corresponding models. 
Moreover, one can employ the mathematical and computational tools developed for statistical and quantum physics for machine learning. For example, mean-field theory and Markov chain Monte Carlo methods originate from statistical physics research are by now standard tools for learning structured probabilistic models~\cite{PGM}. Furthermore, we anticipate that approaches in quantum physics such as tensor networks and quantum algorithms will play an increasingly significant role in generative modeling through the quantum inspired representation of probabilities~\Eq{eq:bornmachine}.

The purpose of this paper is to compare and contrast the \emph{Boltzmann Machines}~(\ref{eq:boltzmannmachine})  and \emph{Born Machines}~(\ref{eq:bornmachine}) approaches for probabilistic modeling, therefore build up a unified view and motivate future studies. Classical and quantum information theories provide crucial guidelines for such comparison. 
Classical information theory lays a common foundation for many problems in machine learning and statistical physics~\cite{mackay2003information, mezard2009information}. 
On the other hand, quantum information theory has played a crucial role in characterizing, modeling and simulating quantum states of matter~\cite{2015arXiv150802595Z}. 
It turns out that many of the physically interesting quantum states only occupy a tiny conner of the Hilbert space, which fulfills the area law of the entanglement entropy~\cite{Eisert:2010hq}. Similar observations were independently made in the machine learning community~\cite{Goodfellow-et-al-2016-Book}, that the images encountered in machine learning applications occupy a negligible proportion of the volume of all possible images. By modeling the probability distribution of classical dataset in terms of the quantum states (\ref{eq:bornmachine}),  insights for modeling quantum states~\cite{2015arXiv150802595Z, Eisert:2010hq} can be transferred into generative modeling of  classical data. 
 
The organization of this paper is as follows. Section~\ref{sec:dataset} defines the complexity of a dataset from the classical and quantum information theoretical perspectives. Section~\ref{sec:RBM} discusses the implication of the information theoretic considerations on the probabilistic modeling using the restricted Boltzmann machines. Section~\ref{sec:experiment} carried out numerical experiments on the standard MNIST dataset to support our claims. Finally, Section~\ref{sec:discussion} summarizes our main points and outlook for future directions.

\section{Complexity of Dataset: Classical Mutual Information and Quantum Entanglement Entropy
\label{sec:dataset}} 

Modeling data probability using an energy based model (\ref{eq:boltzmannmachine}) calls for a classical information theoretical analysis. Mutual information (MI) is a fundamental information theoretical concept which quantifies the complexity of probability distribution $\pi(\mathbf{v})$ associated with the dataset. Assuming $\mathbf{x}\in \mathcal{X}$ and $\mathbf{y}\in\mathcal{Y} $ are two subset of the variables and  $\mathbf{v}=\mathbf{x}\cup\mathbf{y}$, their marginal probability distributions are $\pi(\mathbf{x}) =\sum_{\mathbf{y}\in \mathcal{Y}} \pi(\mathbf{x},\mathbf{y})$, and $\pi(\mathbf{y}) =\sum_{\mathbf{x}\in \mathcal{X}} \pi(\mathbf{x},\mathbf{y})$ respectively.  
The MI reads 
\begin{equation}
  I (\mathcal{X}:\mathcal{Y})  =  \sum_{\mathbf{x}\in \mathcal{X}, \mathbf{y}\in \mathcal{Y}} \pi (\mathbf{x}, \mathbf{y}) \ln\left[ \frac{ \pi(\mathbf{x}, \mathbf{y})}{ \pi (\mathbf{x}) \pi (\mathbf{y})} \right].
 \label{eq:MI}
\end{equation}
The MI measures the amount of information shared between the two sets of variables. MI is zero only for independent variables. In this sense, the MI is a stronger criterion than the correlation of variables since having zero correlation does not necessarily imply vanishing MI. The MI can be used as the objective functions in machine learning applications~\cite{Linsker:1988:SPN:47861.47869, bell1995information, alemi2016deep}. Here we adopt a different point view, which treats MI as a complexity measure of the dataset to be modeled.

\begin{figure}
\begin{center}
\includegraphics[width=\columnwidth]{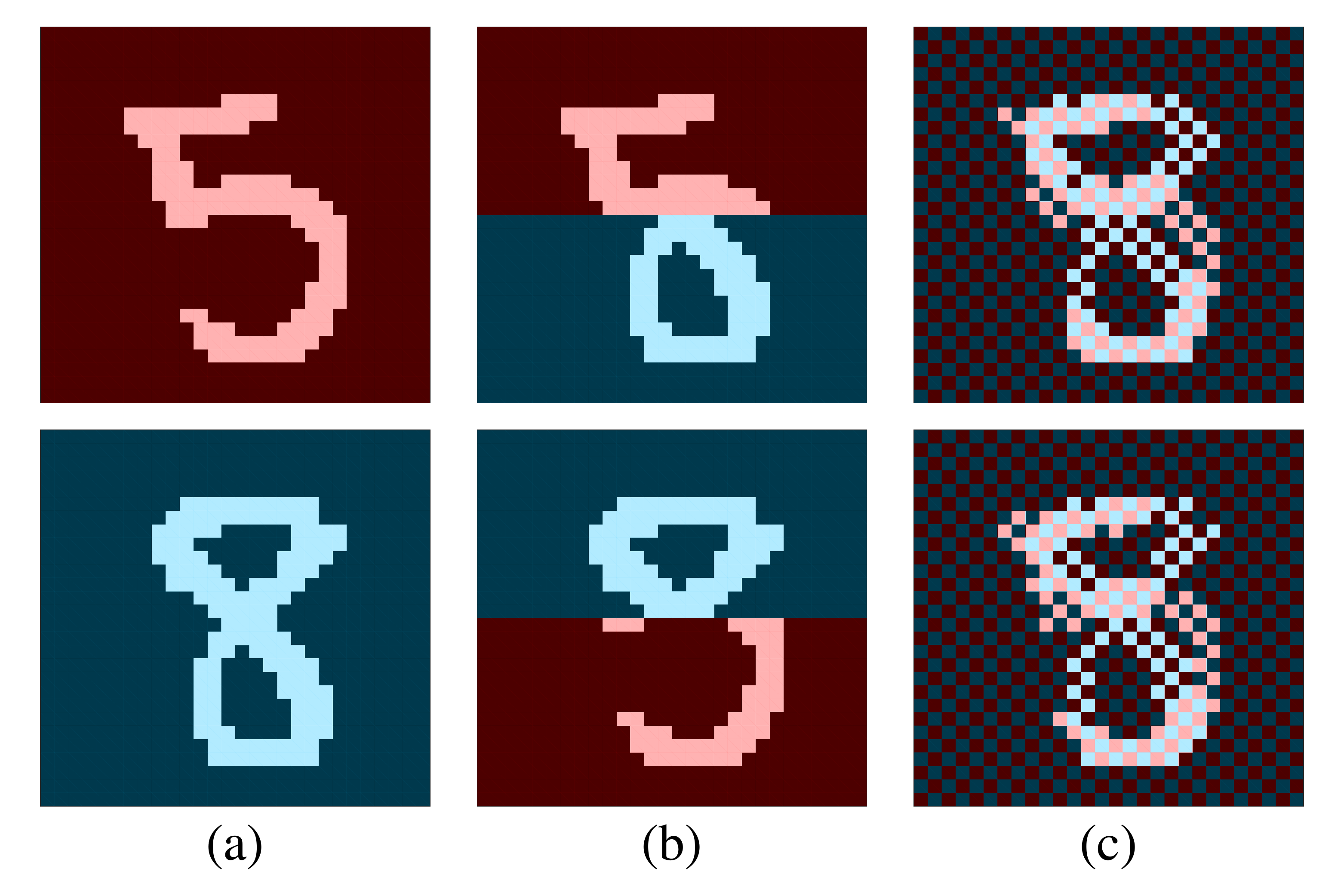}
\caption{Illustration of the swap operation in \Eq{eq:MISWAP} and \Eq{eq:S2SWAP} using handwritten digits from the MNIST dataset. (a) Two original images. (b) Swapped images for up/down bipartition. (c) Swapped images for checkerboard bipartition of the pixels. The blue and red colors indicate the regions of the bipartition $\mathcal{X}$ and $\mathcal{Y}$ respectively.}
\label{fig:swap_figure}
\end{center}
\end{figure}

On the other hand, if we view the target dataset as snapshots of the same quantum state collapsed on a fixed basis (\ref{eq:bornmachine}), it is natural to measure its complexity using the second R\'enyi entanglement entropy
\begin{equation}
S^\mathrm{R} = - \ln \Tr(\rho^{2}_{\mathcal{X}}), 
\label{eq:Salpha}
\end{equation}
where $(\rho_{\mathcal{X}})_{\mathbf{x}, \mathbf{x}^{\prime}} = \sum_{\mathbf{y}\in \mathcal{Y}} \Psi(\mathbf{x}, \mathbf{y}) \Psi(\mathbf{x}^{\prime}, \mathbf{y})$ is the reduced density matrix, and $\Psi(\mathbf{v}=\mathbf{x}\cup\mathbf{y})$ is the probability amplitude associated with the probability, such that $p(\mathbf{v})$ in \Eq{eq:bornmachine} approaches to the data probability distribution $\pi(\mathbf{v})$. 
The second R\'enyi entanglement entropy is a lower bound of the von Neumann entanglement entropy $S^\mathrm{vN} = -\Tr [ \rho_{\mathcal{X}} \ln (\rho_\mathcal{X})]$.

To reveal connection of the classical and quantum information theoretical measures, we write the MI as
\begin{equation}
I(\mathcal{X}:\mathcal{Y}) = - \left\langle  \ln \left\langle
\frac{{\pi}(\mathbf{x},\mathbf{y^{\prime}}){\pi}({\mathbf{x^{\prime}},\mathbf{y}})}
{{\pi}(\mathbf{x^{\prime}},\mathbf{y^{\prime}}){\pi}({\mathbf{x},\mathbf{y})}} \right\rangle_{\mathbf{x}^{\prime},\mathbf{y}^{\prime}}\right\rangle_{\mathbf{x},\mathbf{y}}, 
\label{eq:MISWAP}
\end{equation}
and the second R\'enyi entropy as
\begin{equation}
S^\mathrm{R} = 
 -  \ln  \left\langle \left\langle
\frac{{\Psi}(\mathbf{x},\mathbf{y^{\prime}}){\Psi}({\mathbf{x^{\prime}},\mathbf{y}})}
{{\Psi}(\mathbf{x^{\prime}},\mathbf{y^{\prime}}){\Psi}({\mathbf{x},\mathbf{y})}} \right\rangle_{\mathbf{x}^{\prime},\mathbf{y}^{\prime}}\right\rangle_{\mathbf{x},\mathbf{y}}, 
\label{eq:S2SWAP}
\end{equation}
where the expected value $\langle \cdots \rangle_{\mathbf{x},\mathbf{y}}$ is with respect to the dataset probability $\pi(\mathbf{x},\mathbf{y})$.  

There are apparent similarities between Eqs.~(\ref{eq:MISWAP}) and (\ref{eq:S2SWAP}). Both equations contain swap ratios of probability or probability amplitude~\cite{PhysRevLett.104.157201, PhysRevLett.107.067202}. To illustrated the effect of the swap ratio, Figure~\ref{fig:swap_figure}(a) shows two samples from the MNIST data set [$(\mathbf{x}, \mathbf{y})$ and  $(\mathbf{x}^{\prime}, \mathbf{y}^{\prime})$] and  Fig.~\ref{fig:swap_figure}(b,c) show the corresponding swapped images [$(\mathbf{x}^{\prime}, \mathbf{y})$ and  $(\mathbf{x}, \mathbf{y}^{\prime})$] for up/down and checkerboard bipartitions. The ratio in \Eq{eq:MISWAP} and \Eq{eq:S2SWAP} would be smaller if the swapped images are less likely to appear in the original dataset $\pi(\mathbf{v})$, therefore makes larger contribution to the mutual information or the entanglement entropy. Reference ~\cite{deepandcheaplearn} argues that the dominant correlations in the
natural datasets encountered in physics and machine learning applications are the local ones due to the physical law of the nature. Therefore, it is natural to expect that the checkerboard bipartition [Fig.~\ref{fig:swap_figure}(c)] has higher MI and entanglement entropy compared to the up/down bipartition [Fig.~\ref{fig:swap_figure}(b)] because of strong local correlations between nearby pixels of natural images. Similar discussions on the information measures of different bipartitions were also considered in machine learning~\cite{1704.01552} and in quantum physics~\cite{PhysRevLett.113.106801, PhysRevB.90.075151} studies.

The formal similarity between \Eq{eq:MISWAP} and \Eq{eq:S2SWAP} underlines the analogy between modeling classical data and modeling quantum states~\cite{MPSSL, RBMvsTN, 1704.01552, 2017arXiv170901662H, Liu2017, 2017arXiv171005520Z, 2017arXiv171010248P, 1711.02038}. 
Quantum entanglement entropy is not merely a ``metaphorical vehicle'' to measure the complexity of classical dataset, but is also of practical relevance  if one models the data using the quantum approach~\Eq{eq:bornmachine}. Since the general theories about the entanglement entropy scaling for various quantum states~\cite{Eisert:2010hq} are very instructive for estimating required resources to model the target quantum states, developing of similar theory for typical datasets in machine learning would be very helpful for selecting generative models.

There are nevertheless differences in the two information measures \Eq{eq:MISWAP} and \Eq{eq:S2SWAP}. First, the swap operation in \Eq{eq:MISWAP} is defined for the probability density other than the quantum wavefunction. The probability amplitude may contain phase information which is however irrelevant to probabilistic modeling of the dataset~\cite{2017arXiv170901662H}. Second, the logarithmic functions is sandwiched between two expectations in \Eq{eq:MISWAP}, 
which hiders direct Monte Carlo estimate of the MI similar to the R\'enyi entanglement entropy~\cite{PhysRevLett.104.157201, PhysRevLett.107.067202}. To circumvent this difficulty one may consider to compute alternative quantities such as the R\'enyi mutual information~\cite{PhysRevB.87.195134}.

\section{Probabilistic modeling using Restricted Boltzmann Machine~\label{sec:RBM}}

\begin{figure}
\begin{center}
\includegraphics[width=\columnwidth]{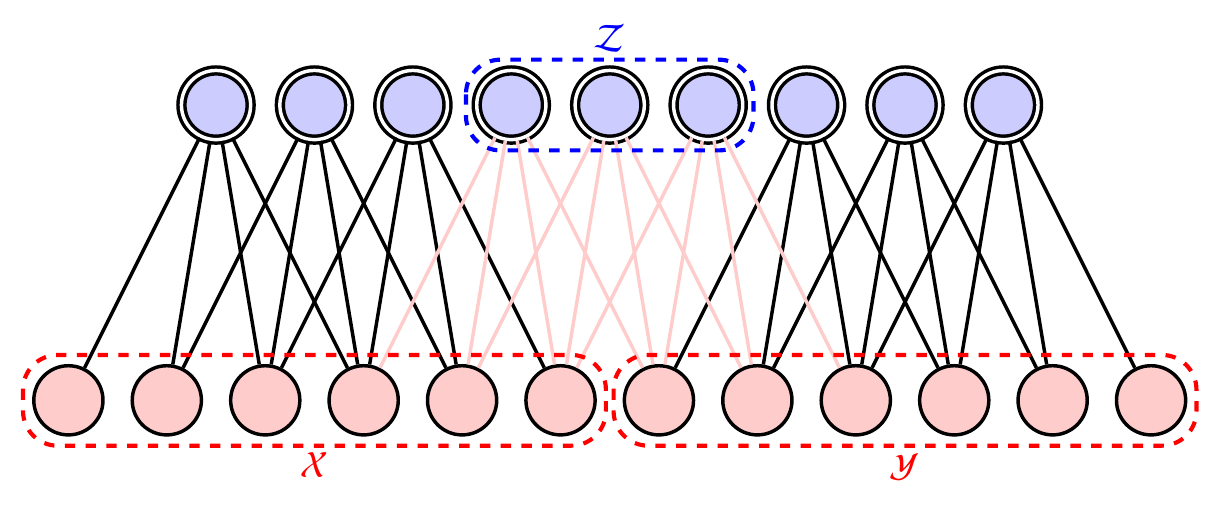}
\caption{A restricted Boltzmann machine consists of visible neurons (red) and hidden neurons (blue with double line) coupled together. The two sets of visible variables $\mathcal{X}$ and $\mathcal{Y}$ are independent once the hidden variables in $\mathcal{Z}$ are given. The red lines are the connections that mediate the interactions between $\mathcal{X}$ and $\mathcal{Y}$ via $\mathcal{Z}$. 
}
\label{fig:rbmxyz}
\end{center}
\end{figure}

As a concrete example, we consider the restricted Boltzmann Machines (RBM)~\cite{Smolensky:1986va} for probabilistic modeling. RBM is a prominent approach for generative modeling with deep connections to statistical physics. 
It has also played an important role in the recent resurgence of deep learning~\cite{HinSal06, Hinton06afast}. Recently, the RBMs have attracted heated attentions in the quantum many-body physics community. Viewed as a variational ansatz for quantum states~\cite{Carleo:2016vp, 2017arXiv170906475N}, the representational power of RBM was investigated from a quantum entanglement~\cite{2017arXiv170104844D} and computational complexity theory~\cite{2017arXiv170105039G} perspectives. Moreover, its connection to the tensor network states was explored extensively~\cite{RBMvsTN, Huang2017,Clark2017,Glasser2017, Kaubruegger2017}.
Besides representing quantum states, RBMs also find applications in identifying order parameters, quantum error correction and accelerating Monte Carlo simulations~\cite{Torlai:2016bm, Huang:2016tf, PhysRevLett.119.030501, 2017arXiv170208586W, 2017arXiv170804622M, 2017arXiv170902597R}. The later applications adopted the conventional usage of the RBMs, i.e., modeling probability density of observed data. 

Conventionally, the RBM models probability distribution of data via an energy-based model with hidden units. By tracing out the hidden variables, the RBM represents a probability distribution of the visible variables. RBM can in principle approximate any probability density by using a sufficiently large number of hidden units~\cite{Freund:1994tu, LeRoux:2008ex, Montufar, montufar2016hierarchical, NIPS2011_4380}. However, one should note that these theorems mostly concern about the worst cases and do not take into account of typical distributions of interests. It is thus crucial to exploit the inductive bias of the RBM in terms of the information measures and match them to the characteristics of the target dataset. To do this, we define the mutual information $I_\mathrm{RBM}$ and entanglement entropy $S^{\mathrm{R}(\mathrm{vN})}_\mathrm{RBM}$ of the RBM analogously to \Eq{eq:MI} and \Eq{eq:Salpha}, except that we now use the probability density $p(\mathbf{v})$ and the corresponding probability amplitude of the RBM.

Given an RBM architecture, one can identify two set of visible variables $\mathcal{X}, \mathcal{Y}$ are connected via a minimal set of hidden variables $\mathcal{Z}$, see Fig.~\ref{fig:rbmxyz}. 
The variables $\mathcal{X}$ and $\mathcal{Y}$ are independent once all the values of $\mathcal{Z}$ are given. This conditional independence property is denoted symbolically as $
\mathcal{X} \bot \mathcal{Y} | \mathcal{Z}
$ in the probabilistic graphical model notation~\cite{PGM}. The MI between the regions $\mathcal{X}$ and $\mathcal{Y}$ can be captured by the RBM is bounded by the size of the intermediate region

\begin{align}
I_\mathrm{RBM}(\mathcal{X}:\mathcal{Y})  \le  
I_\mathrm{RBM} (\mathcal{X}:\mathcal{Z}) 
 \le  
|\mathcal{Z}| \ln 2, 
\label{eq:MIbound}
\end{align}
where $|\mathcal{Z}|$ denotes the number of hidden units in the set $\mathcal{Z}$. The factor $\ln 2$ is due to we consider binary data in this paper. 
The first inequality follows directly from the \emph{data-processing inequality}~\cite{cover2012elements}, which states that the information can not be increased through a random channel. Alternatively, one can show that $I_\mathrm{RBM}(\mathcal{X}:\mathcal{Y})  \le  I_\mathrm{RBM} (\mathcal{X}: \mathcal{Y}\cup \mathcal{Z}) $ using the strong subadditivity property of the MI~\cite{2016arXiv160407450P} and note that $I_\mathrm{RBM} (\mathcal{X}: \mathcal{Y}\cup \mathcal{Z})= I_\mathrm{RBM} (\mathcal{X}:\mathcal{Z}) $~\cite{PhysRevLett.100.070502}. The second inequality in (\ref{eq:MIbound}) uses the fact that mutual information is bounded by the size of the subsystem. 
We note that the mutual information of target data is used for structural learning of fully visible probabilistic graphical model of tree structures 
~\cite{chow1968approximating}. While for RBMs, information theoretical studies have mostly focused on the MI between the visible and hidden variables~\cite{berglund2015measuring, peng2016mutual, koch2017mutual}. 
According to \Eq{eq:MIbound}, one can arrange the hidden neurons of an RBM into a deep architecture, thus to enlarge the size of the intermediate region and increase the  expressibility of the information measures. This motivates the deep Boltzmann Machines~\cite{salakhutdinov2009deep} for more challenging classical dataset with even larger mutual information.

On the other hand, one can repurpose the RBM to represent the quantum state~\cite{Carleo:2016vp}, i.e., the probability amplitude shown in \Eq{eq:bornmachine}. 
In terms of the entanglement entropy, the representational power of RBM is also limited by its connectivity~\cite{2017arXiv170104844D, 2017arXiv170105039G, RBMvsTN, Huang2017}, 

\begin{equation}
S^\mathrm{R}_{\mathrm{RBM}}  \le S^\mathrm{vN}_{\mathrm{RBM}} \le |\mathcal{Z}| \ln 2. 
\label{eq:Sbound}
\end{equation}

Equations~(\ref{eq:MIbound}) and ~(\ref{eq:Sbound}) quantify the expressibility of the RBM in terms of information theoretical measures solely by its architecture. For an RBM with dense connection, the region $\mathcal{Z}$ will span to all the hidden units irrespective of the information pattern of the target dataset.  
Information perspective provides a guiding principle for RBM architecture design conditioned on the typical information pattern of the target dataset. 
Equation (\ref{eq:MIbound}) shows that two sets of visible variables of an RBM should connect to at least $I(\mathcal{X}:\mathcal{Y}))/\ln 2$ hidden neurons to adequately capture the MI of the dataset. We anticipate that the MI of natural images and physical model should be much smaller than the maximum value $\min\left(|\mathcal{X}|, |\mathcal{Y}|\right)\ln 2$ due to physical natural of the probability distributions. 
The connectivity of the RBM puts a constraint on the maximum information can be captured, therefore limits its expressibility. Conversely, this also provides an inductive bias towards what can be easily learned.  
Interpreting the generative modeling in terms of capturing the MI or entanglement of the target dataset shed new light on the learning process.

An important question relevant to quantum machine learning is to identify realistic datasets which are significantly easier to model in the quantum approach than the classical approach~\cite{Perdomo-Ortiz2017}. In light of the above discussion, one is inclined to look for those cases in datasets where the entanglement entropy lower bound \Eq{eq:Sbound} is much smaller than the classical mutual information lower bound \Eq{eq:MIbound}. We verified numerically that in general there is no definite inequality between Eqs.~(\ref{eq:MISWAP}) and ~(\ref{eq:S2SWAP}). Therefore, it would be interesting to construct explicit examples where the quantum approach requires fewer resources.

One should nevertheless be careful when drawing the analogy between modeling quantum states and classical datasets.  For example, it is sometimes argued that one  needs deep neural nets to model classical dataset with critical correlations in analog to critical quantum systems which can only be captured by hierarchical tensor networks~\cite{PhysRevLett.101.110501}. However, the scaling behavior of the mutual information of a critical classical system is different from the entanglement entropy of a critical quantum system. The MI of statistical physics model with short-range interactions scales only with the boundary size between subsystems~\cite{PhysRevLett.100.070502}, which holds irrespective whether the system is critical or not.  As a concrete example, the critical Ising model only requires a shallow RBM to be modeled exactly~\cite{RBMvsTN}, which is in line with the area law scaling of its mutual information~\cite{1742-5468-2011-10-P10011, PhysRevE.87.022128}. Reference~\cite{2017arXiv170804622M} also shows that deep architectures do not seem to exhibit advantages for modeling the critical Ising data.

Reference~\cite{2017arXiv170305334T} use an RBM to model the probability of a quantum state on a fixed basis for quantum state tomography. Since the approach corresponds to the \Eq{eq:boltzmannmachine}, the required resources are determined by the Shannon mutual information of quantum states~\cite{PhysRevLett.111.017201, PhysRevB.90.045424}, instead of the entanglement entropy. The two entropies exhibit similar scaling behavior for the examples discussed in Refs.~\cite{PhysRevLett.111.017201, PhysRevB.90.045424}. However, in general this may not be the case. Thus, it remains open to see whether is it advantageous to use an RBM to model the probability or using a complex valued RBM to model the quantum state directly. 

\begin{figure}
\begin{center}
\includegraphics[width=\columnwidth]{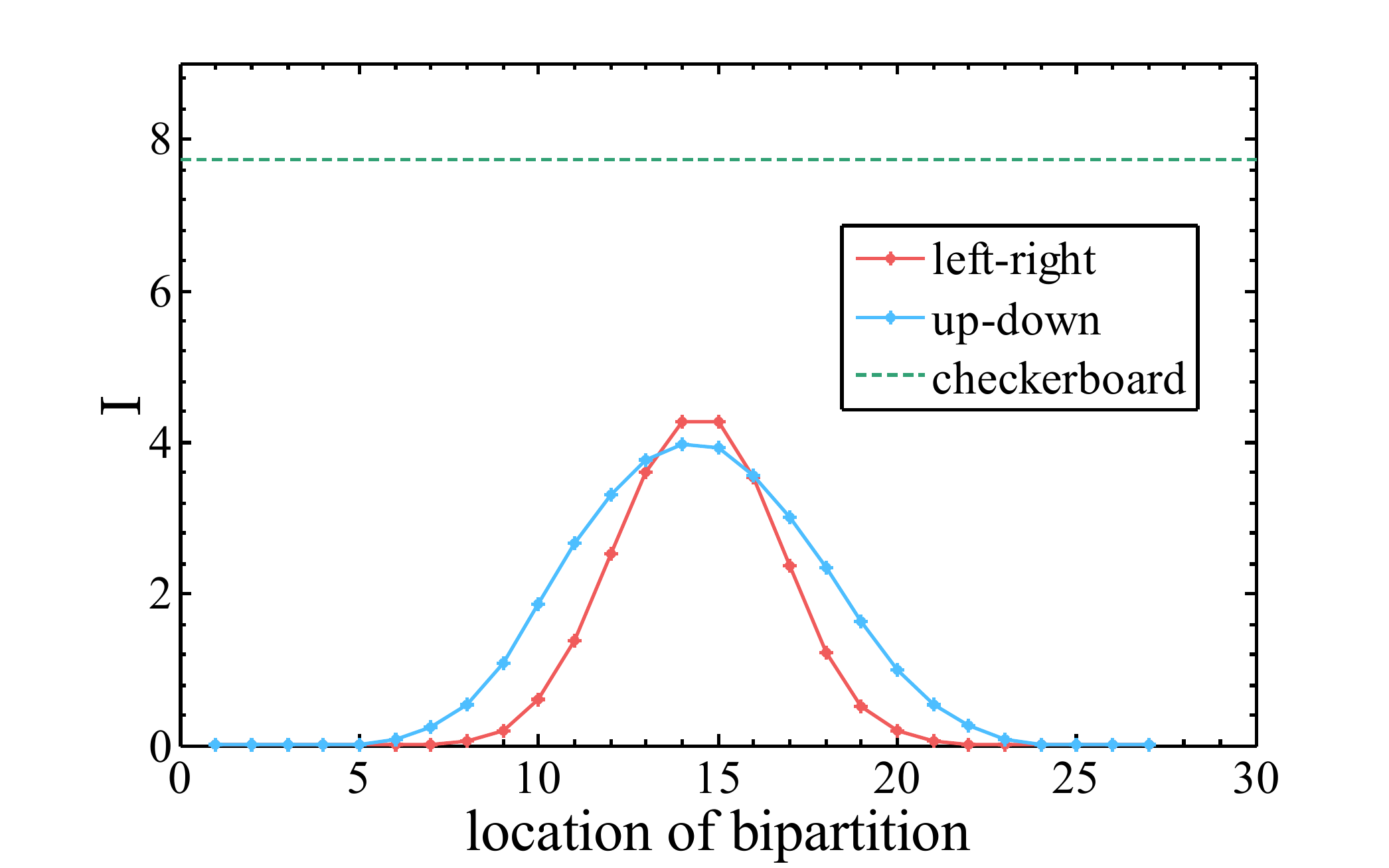}
\caption{Mutual information \Eq{eq:MI} of the MNIST dataset for various bipartitions of the images.
}
\label{fig:KSG}
\end{center}
\end{figure}

\section{Mutual information of MNIST dataset and its implication to RBM architecture design\label{sec:experiment}}

We consider modeling the MNIST dataset using RBMs by exploiting the mutual information patter of the target dataset. First, we employ the approach of~\cite{PhysRevE.69.066138} to estimate the MI of the MNIST dataset. The approach is based on nearest neighbor estimate of the Shannon entropy which is widely adopted in the statistics literature. 
Figure~\ref{fig:KSG} shows how the MI increases as one cuts into the center of the image. The vertical and horizontal bipartition of the images exhibit similar behavior of the MI. The MI between the margin and the remaining part of the image is zero since the margin of the MNIST image is always fixed. MI of the checkerboard bipartitions such as Fig.~\ref{fig:swap_figure}(c) is significantly higher than the left/right or up/down bipartitions. This suggests that introducing hidden units which couple to the nearby pixels of the MNIST images are more efficient in capturing the MI of the MNIST dataset. Although this appears to be quite obvious to the MNIST dataset, similar analysis would be instructive for less familiar datasets. 
As a side note, the MI estimator~\cite{PhysRevE.69.066138} is only approximate, especially for highly dependent variables~\cite{Gao2014}. It is generally a difficult task to  estimate the MI of image dataset rigorously. On the other hand, estimating the entanglement entropy~(\ref{eq:S2SWAP}) is feasible by using tensor network~\cite{MPSSL, 2017arXiv170901662H, Liu2017} or Monte Carlo approaches~\cite{PhysRevLett.104.157201, PhysRevLett.107.067202}. 

\begin{figure}
\begin{center}
\includegraphics[width=\columnwidth]{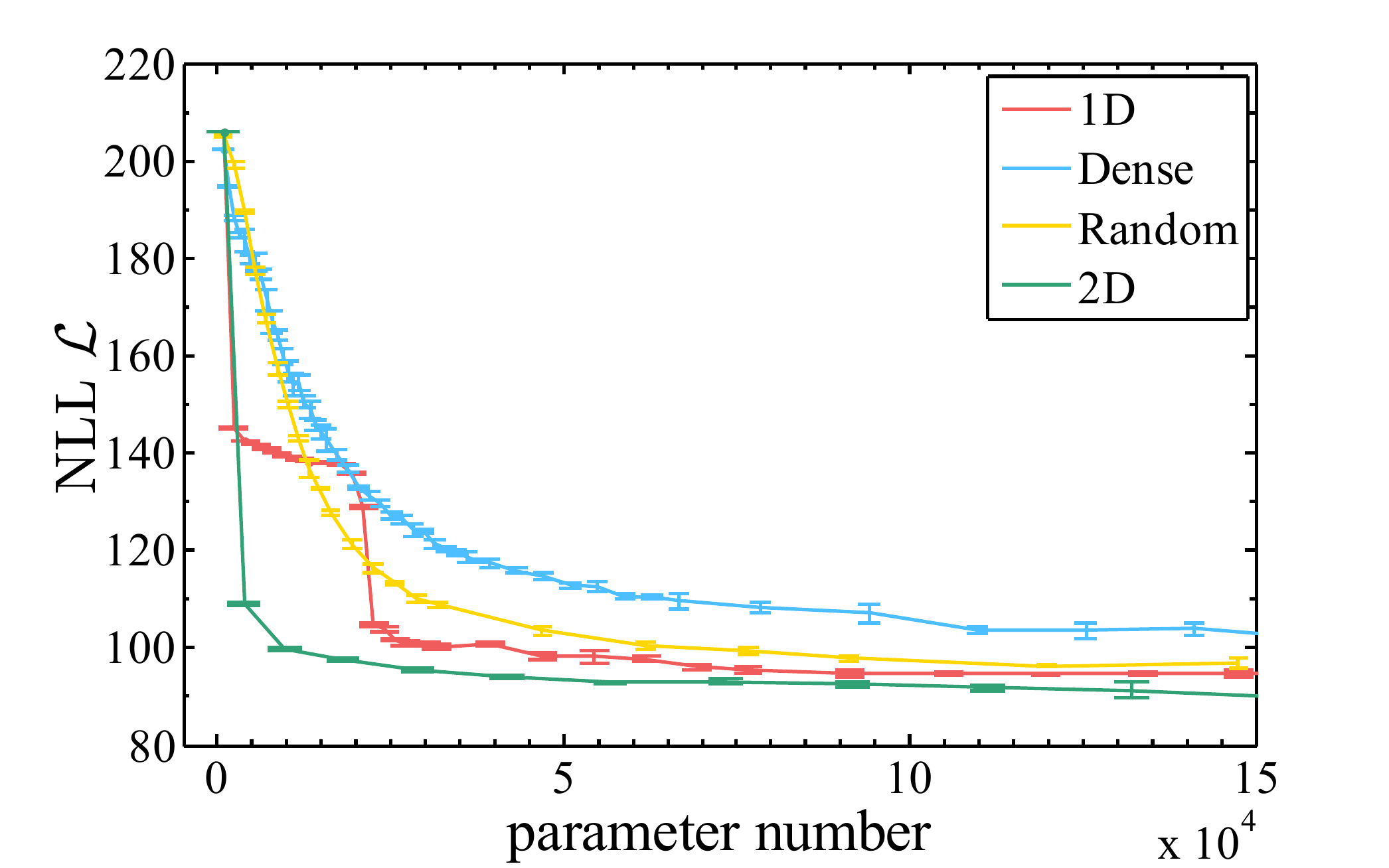}
\caption{Negative loglikelihood for various RBMs architectures plotted against the number of parameters in the model. 
}
\label{fig:parameter_num_compare}
\end{center}
\end{figure}

The distribution of mutual information of the MNIST dataset suggests that local connections are more important to capture the mutual information of the dataset. We verify this by training RBMs with the same number of parameter but with different connection architectures and number of hidden neurons. Dense connection means that the visible and hidden units of the RBM are fully connected. Random means that we randomly connect the visible and hidden neurons. 1D means each hidden neurons of the RBM connected only a fragment of the entire image vector, see Fig.~\ref{fig:rbmxyz}. 2D means each hidden neurons connected a small window of a 2D image. The goodness of the learning is measured by the negative log-likelihood (NLL) evaluated on the test dataset $\mathcal{L} = -\frac{1}{|\mathcal{D}|}\sum_{\mathbf{v}\in \mathcal{D}} \ln\left[{p(\mathbf{v})}\right]$, 
where $|\mathcal{D}|$ is the size of the test set. To estimate the partition function involved in the NLL computation we employ the Annealed Importance Sampling approach~\cite{neal2001annealed, salakhutdinov2008}. The estimated NLL provides an up bound of the entropy of the dataset, which also bounds the MI between two arbitrary division of the variables, i.e. $\mathcal{L} \ge I(\mathcal{X}:\mathcal{Y})$.

One can clearly see in Fig.~\ref{fig:parameter_num_compare} that the RBM  respects the 2D nature of the images with local connections reaches the lowest NLL quickly with the least number of parameters. While the NLL of the RBM with 1D connections exhibits an abruptly drop when the two nearby pixels from different row are connected. The RBM with dense connections performs even worse than the random connections given the same number of parameters. 
Our results are in consistent with the previous experiments on RBMs with sparse connections. Reference~\cite{2013arXiv1312.5258D} shows the neural network works fine even with $80\%$ randomly dropped out. Reference~\cite{Mocanu2016} also proposed a sparsely connected RBM with small-world network structure and found that it performs well compared with a densely connected RBM.

One can see that by exploiting the mutual information pattern of the target dataset in the RBM structure design one can greatly enhance the representing and learning efficiency. 
Since RBMs with local and sparse connections exhibit close connections to the tensor networks~\cite{RBMvsTN}, the above experiments also support the applications of tensor network states in machine learning problems~\cite{MPSSL, 2017arXiv170901662H, Liu2017}. In those applications, it is more natural to adopt \Eq{eq:bornmachine}) and the associated quantum information perspectives.

\section{Summary \label{sec:discussion}}

In summary, revealing the similarity of the two information theoretical measures Eqs.~(\ref{eq:MISWAP},~\ref{eq:S2SWAP}) suggests that the statistical physics and quantum physics inspired approaches for generative modeling~ \Eq{eq:boltzmannmachine} and \Eq{eq:bornmachine} have similar inductive biases. Therefore, successful wavefunction representations in quantum physics have  the potential to be good generative models for machine learning, and vice versa. 

Classical and quantum information theories shed light on the expressibility and architecture design of generative models. Our discussions and numerical experiments suggest that it is rewarding to design architectures which take into information pattern of the target dataset. In particular, imposing locality greatly increases the learning efficiency by exploiting the mutual information structure of typical dataset. This is akin to the success of the convolutional neural network structure for discriminative tasks.

Besides the expressibility issue discussed in the paper, learning and sampling of the energy-based models can be slow due to the intractable partition functions. Conventionally, this is solved by using Markov chain Monte Carlo sampling or mean-field theory approaches. 
The quantum representation offers an alternative solution to these problems. For example, modeling the probability amplitude as matrix product states or tree tensor networks offer advantages in efficient learning and sampling~\cite{MPSSL, 2017arXiv170901662H, Liu2017}. Moreover, representing the probability distribution using a quantum state~\cite{1711.02038, Farhi2016} obviously permits efficient sample generation by simply performing measurement to the quantum state. 

In this paper we discussed the energy-based models and quantum state representations for probabilistic modeling of classical data. Nevertheless, a combined approach with a ``quantum statistical model'' is also possible, in which one models the classical probability density using mixed quantum states. In this respect, the quantum Boltzmann machines~\cite{2016arXiv160102036A,Kieferova2016, 2016arXiv160902542B} can be viewed as an example. Finally, this paper focused on the modeling the probability of data without labels. In a general setting, one could also model the joint probability distribution of the data and label. In this case, one can generate samples conditioned on the class label and elaborate on the entanglement entropy of each class individually~\cite{Liu2017}.

\begin{acknowledgments}
We thank E. Miles Stoudenmire, Qianyuan Tang, Cheng Peng, Zhao-Yu Han, Jun Wang, Pan Zhang for inspiring discussions and collaborations. 
S.C. and J.C. are supported by the National R\&D Program of China (Grant No. 2017YFA0302901) and the National Natural Science Foundation of China (Grants No. 11190024 and No. 11474331).
L.W. is supported by the Ministry of Science and Technology of China under the Grant No. 2016YFA0300603 
and National Natural Science Foundation of China under the Grant No. 11774398. 
\end{acknowledgments}

\bibliography{MI_RBM}

\end{document}